\documentclass[prl,twocolumn,groupedaddress,superscriptaddress]{revtex4}
\usepackage{graphicx}
\usepackage{amssymb}
\usepackage{marvosym}
\usepackage{latexsym}
\usepackage[usenames]{color}
\usepackage{float}
\usepackage{array}

\begin{document}

\title{Linking Dynamic and Thermodynamic Properties of Cuprates; an ARPES
study of (CaLa)(BaLa)$_{2}$Cu$_{3}$O$_{y}$}
\author{Gil Drachuck}
\affiliation{Department of Physics, Technion - Israel Institute of Technology, Haifa,
32000, Israel}
\author{Elia Razzoli}
\affiliation{Swiss Light Source, Paul Scherrer Institute, CH-5232 Villigen PSI,
Switzerland}
\author{Rinat Ofer}
\affiliation{Department of Physics, Technion - Israel Institute of Technology, Haifa,
32000, Israel}
\author{Galina Bazalitsky}
\affiliation{Department of Physics, Technion - Israel Institute of Technology, Haifa,
32000, Israel}
\author{R. S. Dhaka}
\affiliation{Swiss Light Source, Paul Scherrer Institute, CH-5232 Villigen PSI,
Switzerland}
\author{Amit Kanigel}
\affiliation{Department of Physics, Technion - Israel Institute of Technology, Haifa,
32000, Israel}
\author{Ming Shi}
\affiliation{Swiss Light Source, Paul Scherrer Institute, CH-5232 Villigen PSI,
Switzerland}
\author{Amit Keren}
\affiliation{Department of Physics, Technion - Israel Institute of Technology, Haifa,
32000, Israel}
\date{\today }

\begin{abstract}
We report angle-resolved photoemission spectroscopy (ARPES) on two families
of high temperature superconductors (Ca$_{x}$La$_{1-x}$)(Ba$_{1.75-x}$ La $%
_{0.25+x}$)Cu$_{3}$O$_{y}$ with $x=0.1$ ($T_{c}^{max}=56$~K) and $x=0.4$ ($%
T_{c}^{max}=82$~K). The Fermi surface (FS) is found to be independent of $x$
or $y$, and its size indicates extreme sample-surface overdoping. This
universal FS allowes the comparison of dynamical properties between
superconductors of similar structure and identical doping, but different $%
T_{c}^{max}$. We find that the high-energy ($\left\vert E\right\vert >50$%
~meV) nodal velocity in the $x=0.4$ family is higher than in the $x=0.1$
family. The implied correlation between $T_{c}^{max}$ and the hopping rate $%
t $ supports the notion of kinetic energy driven superconductivity in the
cuprates. We also find that the antinodal gap is higher for the $x=0.4$
family.
\end{abstract}

\maketitle

The recent synthesis of charge compensated (Ca$_{x}$La$_{1-x}$)(Ba$_{1.75-x}$%
La$_{0.25+x}$)Cu$_{3}$O$_{y}$ (CLBLCO) single crystals facilitates an
investigation of the relationship between their dynamical properties, such
as the electronic dispersion relation $E(\mathbf{k})$ and their
thermodynamic property $T_{c}$, while applying subtle crystal structure
changes \cite{Crystal}. Since the valence of Ca and Ba is equal, $x$ has a
minute effect on crystal structure but a large effect on $T_{c}$.
Therefore, CLBLCO allows experiments where the correlations between $T_{c}$
and a single parameter are explored. Experiments of such nature
can reveal the mechanism for cuprates' superconductivity. In the present
work, we measure the electron dispersion $E(\mathbf{k})$ of two extreme
samples of CLBLCO crystals, using angle-resolved photoemission spectroscopy
(ARPES), and look for correlations between properties of $E(\mathbf{k})$
and $T_{c}$. In particular, we focus on the nodal velocity. Previously,
similar studies could only be done by comparing cuprates with very different
structures and levels of disorder \cite{EdeggerPRL06}.

\begin{figure*}[tbph]
\begin{center}
\includegraphics[trim=0cm 0cm 0cm 0cm,clip=true,width=15cm]{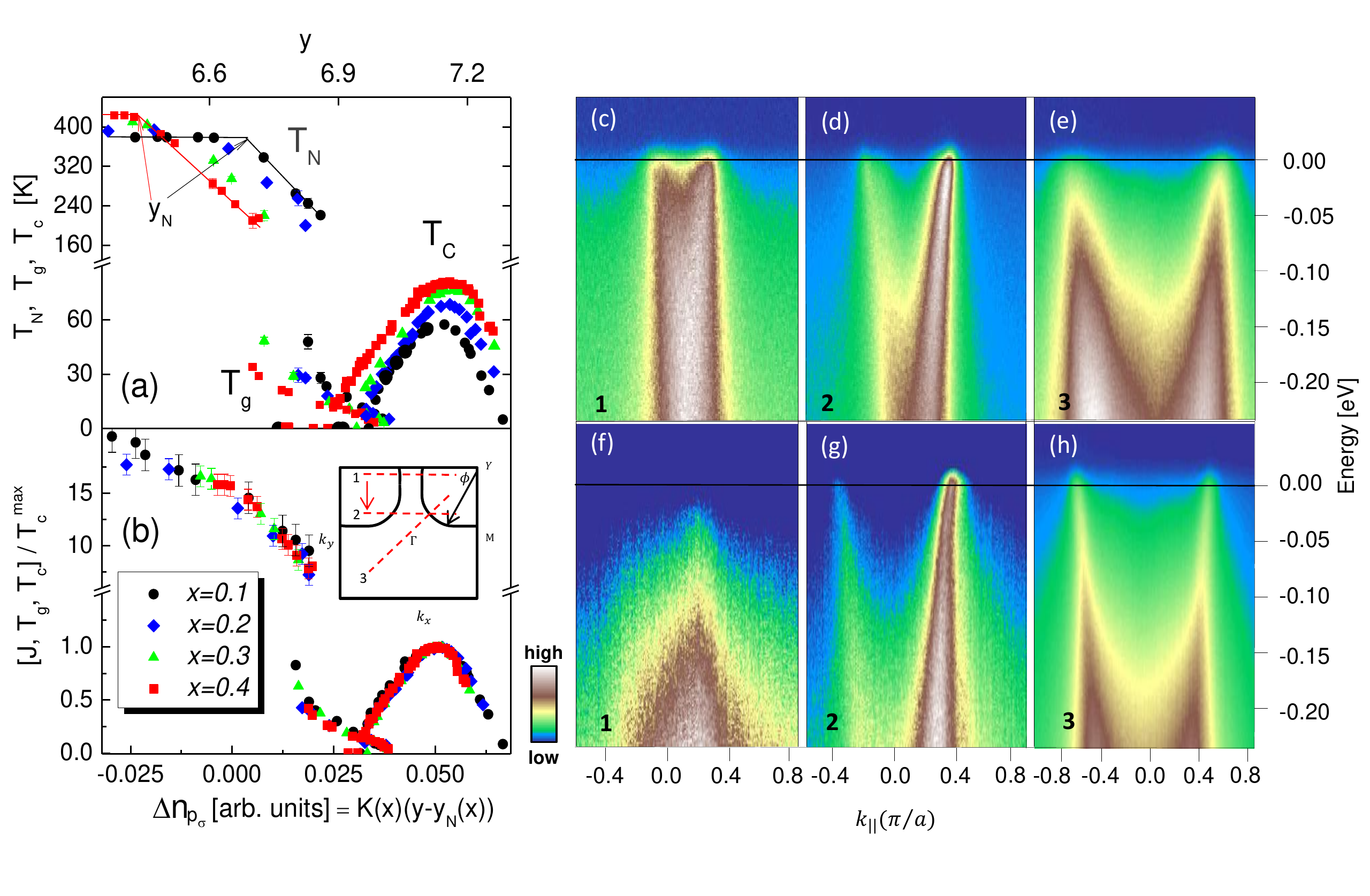}
\end{center}
\caption{(a) The phase diagram of CLBLCO showing the N\'{e}el ($T_{N}$),
glass ($T_{g}$) and superconducting ($T_{c}$) temperatures over the full
doping range for the four families. $y_{N}$ indicates the oxygen level where
$T_{N}$ start to drop. (b) The unified phase diagram of CLBLCO. The critical
temperatures, and $J$ extracted from $T_{N}$, are divided by $T_{c}^{max}$
and plotted as a function of doping variation in the oxygen orbital $\Delta
n_{p_{\protect\sigma }}$. (c)-(e) Raw ARPES date measured on a CLBLCO $x=0.1$
crystal at $T=16$~K. The numbers on the figures correspond to cut
trajectories illustrated in the inset of the phase diagram. (f)-(h) The same
as (c)-(e) but for a sample with $x=0.4$ measured at $T=11$~K.}
\label{Fig1}
\end{figure*}

CLBLCO is similar to YBCO in crystal structure, but has no oxygen chain
ordering and is tetragonal for all $x$ and $y$ \cite{Yaki99}. This
simplifies the ARPES interpretation. While $x$ alters the calcium-to-barium
ratio, the lanthanum content in the chemical formula remains constant. We
define four CLBLCO "families" as samples with different $x$, namely, $%
x=0.1,0.2,0.3,0.4$. The parameter $y$ signifies the oxygen level, which
drives the system between different phases. By varying $x$ and $y$ in
the chemical formula, one can generate phase diagrams that are similar in
shape yet differ in the maximum of $T_{c}$, $T_{g}$, and $T_{N}$, and in the
critical oxygen level at which the nature of the phase diagram changes. The
phase diagram is presented in Fig.~\ref{Fig1}(a) \cite{AmitPRB10}. It is
worth noting that the only structural properties that vary with $x$ or $y$
are the Cu-O-Cu buckling angle, bond length, and CuO$_{2}$ plane
doping efficiency $K(x)$. The crystallographic parameters were measured with
powder neutron diffraction \cite{OferPRB08}. The buckling angle decreases by
$0.5$ degrees as $x$ increases between families. The bond length varies from $%
3.88$~\AA\ for $x=0.4$ to $3.91~$\AA\ at $x=0.1$. The doping efficiency is
determined by in-plane $^{17}$O nuclear quadrupole resonance (NQR) \cite%
{AmitPRB10}. The variation in the number of holes on an oxygen $\Delta
n_{p_{\sigma }}$ is given by $\Delta n_{p_{\sigma }}=K(x)(y-y_{N})$, where $%
y_{N}$ is defined as the doping at which $T_{N}$ starts to drop [see Fig.~%
\ref{Fig1}(a)] \cite{AmitPRB10}.

The super-exchange parameter $J$ \ for each CLBLCO family was previously
determined with muon spin rotation ($\mu $SR) (magnetization) versus
temperature measurements \cite{OferPRB06} and with two magnon Raman
Scattering \cite{Dirk}. Figure~\ref{Fig1}(b) depicts the super-exchange $J$
and glass temperature $T_{g}$ (both from $\mu $SR), and $T_{c}$, all
normalized by $T_{c}^{max}$, as a function of $\Delta n_{p_{\sigma }}$. A
universal phase diagram appears, demonstrating that $T_{c}^{max}$ scales
like $J$ \cite{AmitPRB10}, which implies that $T_{c}^{max}$ is determined by
the overlap of the orbital occupied by electrons on neighboring sites.
Orbital overlaps also determine the hopping parameter $t$, and the scaling
of $T_{c}^{max}$ with $J$ meaning that kinetic energy controls the
superconducting phase transition. However, $J$ is determined in the AFM
phase, which is \textquotedblleft far\textquotedblright , in terms of
doping, from the superconducting phase. A question arises: are the orbital
overlaps important in the superconducting phase as well? In this phase $t$ can be measured. Here, we extract $t$ from $E(\mathbf{k}%
)$ as the velocity in the nodal direction. We find correlations between $%
T_{c}^{max}$ and $t$, and confirm the famous relation $J\propto t^{2}$~\cite%
{Eskes-tJ}. This suggests that the band structure is rigid as a function of
doping, as suggested by recent resonance inelastic x-ray scattering
experiments \cite{LeTaconNP11}. By the same token, we also measure the
antinodal gaps and compare them with Hamiltonian parameters.

The ARPES experiments were performed on the SIS beam-line at the Swiss Light
Source on CLBLCO single crystals. These unique crystals were grown using the
traveling floating zone method. A detailed discussion about growth and
characterization of these crystals is given in~\cite{Crystal}. For this
experiment, samples with $x=0.1$ and $x=0.4$ were used. The samples were
mounted on a copper holder with silver glue to improve electrical
conductivity. The Fermi level and resolution were determined from the
polycrystalline copper sample holder. The samples were cleaved \textit{in
situ} using a glued-on pin at $T=10-20$~K. Circularly polarized light with $%
h\nu =50$~eV was used. The spectra were acquired with a VG Scienta R4000
electron analyzer. Despite a base pressure of $5 \times 10^-11$ torr, the samples' surface life-time was only a few hours and a
high intensity beam was required for quick measurements. As a consequence,
the energy resolution in our experimental conditions was limited to $17-22$
meV.

In Fig.~\ref{Fig1}, we present ARPES data collected from CLBLCO for the
two samples: $x=0.1$ is presented in panels c, d, and e, while $x=0.4$ is
depicted in panels f, g, and h. The data was collected at $T=16$~K and $11$%
~K for the $x=0.1$ and $0.4$ respectively. All spectra are normalized by the
measured detector efficiency. For each sample, intensities along three cuts
are shown. The cuts are illustrated and numbered on the Fermi surface (FS)
drawing in the inset of Fig.~\ref{Fig1}(b). Cuts numbered 1 and 2 are along $%
k_{x}$ ($\Gamma -M$ direction). These cuts allow better sensitivity to the
gap size at the antinode. Cut number 3 is along the diagonal line of the BZ (%
$\Gamma -Y$). In this configuration, a measurement of velocity in the
nodal direction is possible. The number on the bottom of each ARPES panel
indicates the cut from which data are collected.

In Figs.~\ref{Fig1}(c) and \ref{Fig1}(f), spectra near the anti-node are
plotted. While $x=0.1$ shows high intensity spectra up to $E_{f}$ where no
gap is visible, the $x=0.4$ sample shows a depletion of intensity close to $%
E_{f}$, indicating a gap in the spectra at the antinode. For the $x=0.1$,
the gap, if one exists, is smaller than the experimental resolution. In
Figs.~\ref{Fig1}(d) and \ref{Fig1}(g), we plot the intensity closer to the
node. For both the $x=0.1$ and $0.4$ sample, we clearly see the spectra
crossing $E_{f}$ indicating a closed gap in the nodal region. Finally, in
Figs.~\ref{Fig1}(e) and \ref{Fig1}(h), both nodal cuts are seen, and again,
the spectra cross $E_{f}$, indicating an absence of a gap along the Fermi
arc for both samples. The last panels also show a clear dispersion from
which the nodal velocity is extracted.

\begin{figure}[tbph]
\begin{center}
\includegraphics[trim=2.7cm 0cm 0cm 0cm,clip=true, height=8cm]{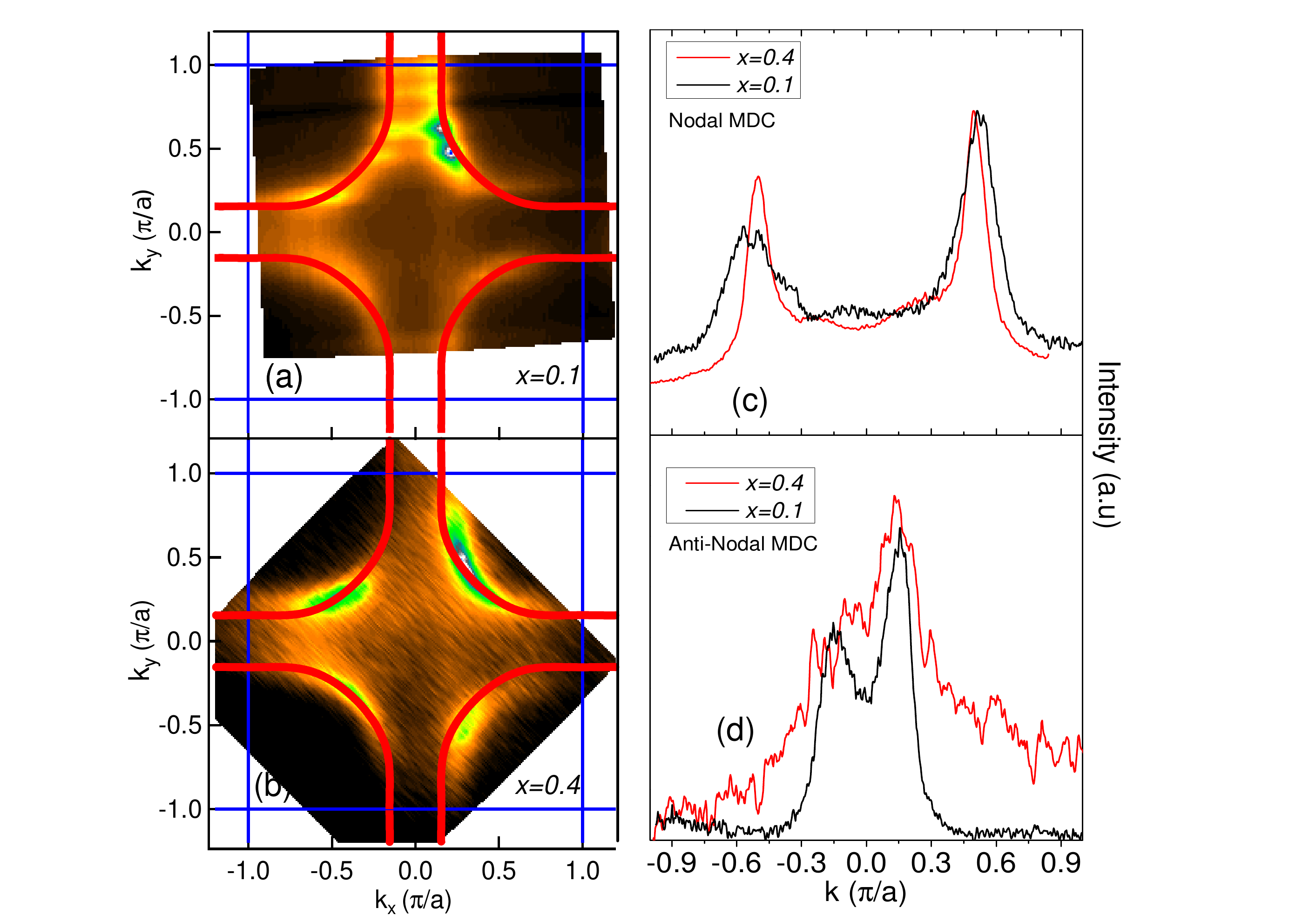}
\end{center}
\caption{(a)-(b) Spectral weight map in $\mathbf{k}$-space at $E_{f}$ (FS)
in CLBLCO $x=0.1$ and $x=0.4$ sample, respectively. The data were obtained
at T=16 K for the $x=0.1$ sample and at T=11 K for the $x=0.4$ sample. Both
samples were prepared with optimal doping and verified with a SQUID
magnetometer. The red curve is the FS of CLBLCO obtained from tight binding
fits to experimental data (see text). (c) MDCs at zero binding energy
along a nodal cut, for the $x=0.1$ (black, Fig.~\protect\ref{Fig1}(e)) and $%
x=0.4$ (red, Fig.~\protect\ref{Fig1}(h)) samples. (d) MDCs at zero binding
energy along the antinodal cut, for the $x=0.1$ (black, Fig.~\protect\ref%
{Fig1}(d)) and $x=0.4$ (red, Fig.~\protect\ref{Fig1}(g)) samples.}
\label{Fig2}
\end{figure}

In Fig.~\ref{Fig2}, we show the FS in the first Brillouin zone (BZ), for the
two CLBLCO samples: $x=0.1$ (Fig.~\ref{Fig2}(a)) and $0.4$ (Fig.~\ref{Fig2}%
(b)). The FS was obtained by integrating 10 meV around the chemical
potential. The ARPES intensity is displayed in a false color scale as a
function of $k_{x}$ and $k_{y}$. By comparing the shape of the FS, we can
see that the $x=0.4$ sample exhibits a Fermi arc structure~\cite%
{kanigel-arcs}, which is typical for an antinodal gap. As for the $x=0.1$
sample, the arc is not present, and we observed strong intensity at the
antinode, comparable to the intensity near the nodal region. Unlike
previously reported FS measurements of YBCO \cite{k-ybco}, there is no
apparent chain-like structure in the CLBLCO FS, as expected. The red line is
a fit to a tight binding (TB) model up to three nearest neighbours
hopping. The fit parameters will be discussed below. The fit for both FSs
gives the same size, as can be seen in Fig.\ref{Fig2}. In fact, the FS of a
variety of samples was measured and found to be identical regardless of
family ($x$) or bulk oxygen level ($y$).

A clearer comparison of the FS size and doping between families can be
obtained by examining the node-to-node and antinodal distances. In Fig~\ref%
{Fig2}(c), we show momentum distribution curves (MDC) at zero binding
energy ($E_{f}$) measured in a nodal cut ("cut 3"), for both samples. The
MDC for $x=0.4$ is sharper than for $x=0.1$, but the peak-to-peak distance
is equal for both MDCs. Similarly, Figure~\ref{Fig2}(d) depicts an MDC
measured in the antinode ("cut 2") at $E_{f}$. Here, the MDC of $x=0.1$ is
clearer than that of $x=0.4$ because of an open gap, but again the peak
separation for both samples is identical.

We suspect that the bulk doping independence of the FS is due to the sample
being cleaved on a charged plane, inducing surface charge reconstruction.
Such behavior was previously reported from measurements of YBCO
\cite{k-ybco}. From the measured nodal peak-to-peak distance as a function
of doping in YBCO described in~\cite{k-ybco}, we can estimate the doping of
our sample, which turns out to be $p=0.23\pm 0.02$. This result is
consistent with calculations based on the FS volume. Thus, we can conclude
that the surface doping level of both samples is equal within the
experimental error, and that the surface doping is on the edge of the
superconducting dome on the overdoped side.

\begin{figure}[b]
\begin{center}
\includegraphics[trim=2cm 0cm 0cm 0cm,clip=true,height=8cm]{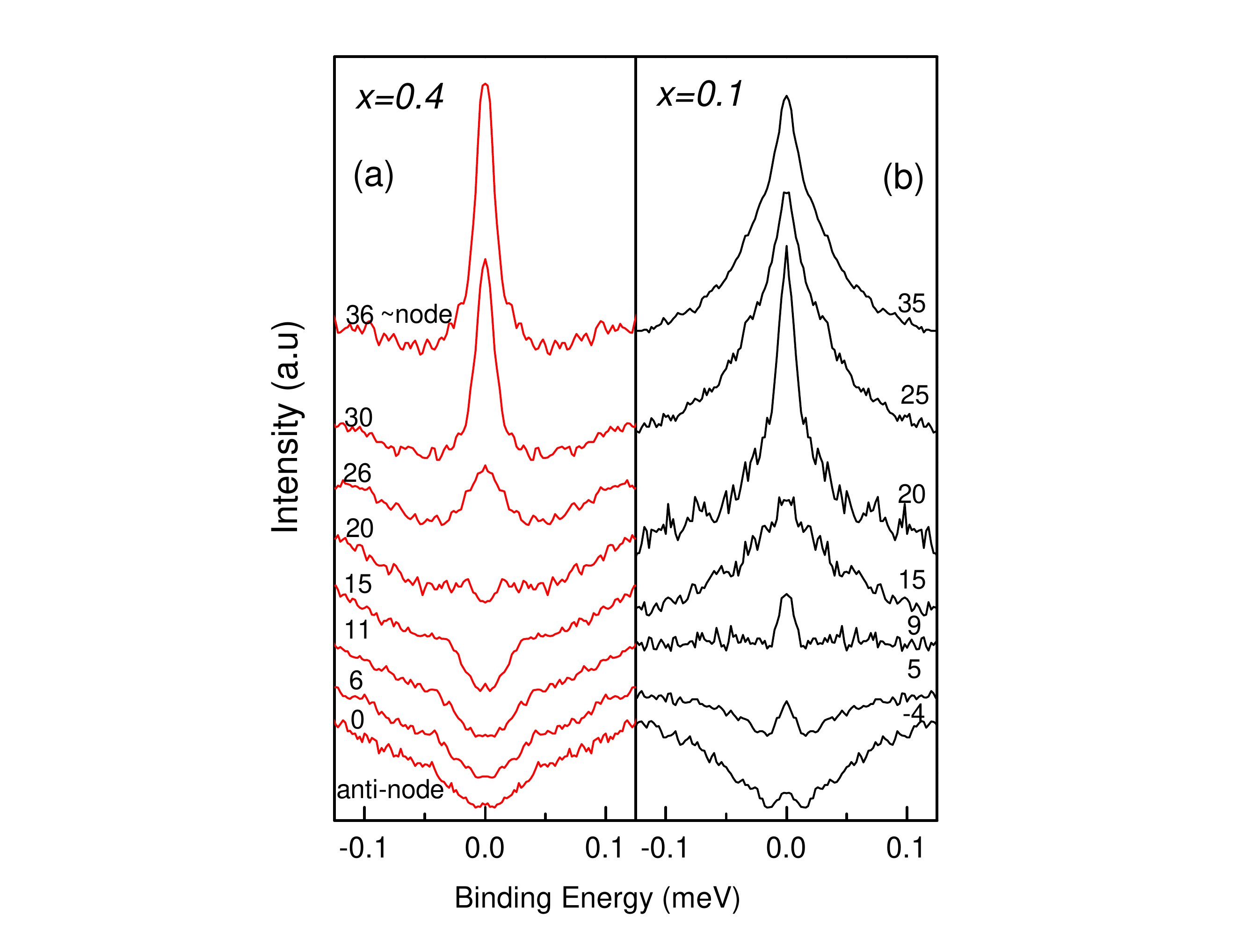}
\end{center}
\caption{(a) Symmetrized EDCs for the $x=0.4$ sample at $k_f$ as a function
of the Fermi surface angle $\protect\phi$, from the node (top) to the
anti-node (bottom). The cuts are measured along the $\Gamma-M$ direction.
Each curve is offset for clarity. (b) the same as (a) but for the $x=0.1$
sample.}
\label{Fig3}
\end{figure}

To investigate the momentum dependence of the gap, we measured the
dispersion along $\Gamma -M$ cuts between \textquotedblleft cut
1\textquotedblright\ and \textquotedblleft cut 2\textquotedblright\ for the $%
x=0.1$ and $x=0.4$ samples at a cold finger temperature of $T=16$ K and $%
T=11 $ K, respectively. In Fig.~\ref{Fig3}, we plot symmetrized EDC's at
$k_{f}$ as a function of FS angle $\phi $ (defined in the inset of Fig~\ref%
{Fig1}). For the $x=0.4$ sample (Fig.~\ref{Fig3}(a)), one can see a
zero-energy intensity peak close to the node ($\phi =36$). In contrast, at
an angle of $\phi =20$ and lower, we observe an opening of a gap, which
grows up to $\Delta _{0}=40$ meV at the antinode ($\phi =0$). The angular
dependence of the gap is shown in Fig.1 of the supplementary material. The
gap value at the antinode is similar to optimally doped Bi2212~\cite%
{gap-doping-bssco,mesot-bi2212gap} and YBCO~\cite{gap-ybco,YBCO-gap-shen}.

For the $x=0.1$ sample [Fig.~\ref{Fig3}(b)], the situation is different.
Close to the node, we observe a strong peak at zero energy ($\phi =35 $). As
we move to the antinode, the intensity at zero energy is partly suppressed,
but unlike the $x=0.4$ sample, there is no full depletion of spectral
density at $E=0$. This indicates that a gap is not present in the $x=0.1$
sample, or that it is smaller than the experimental resolution ($20$ meV). A
closed gap was measured with the same resolution for two more $x=0.1$
samples. Thus, we can safely say that $\Delta _{0}(x=0.4)>\Delta _{0}(x=0.1)$%
.

\begin{figure}[tbph]
\begin{center}
\includegraphics[trim=2cm 0cm 0cm 0cm,clip=true,height=7cm]{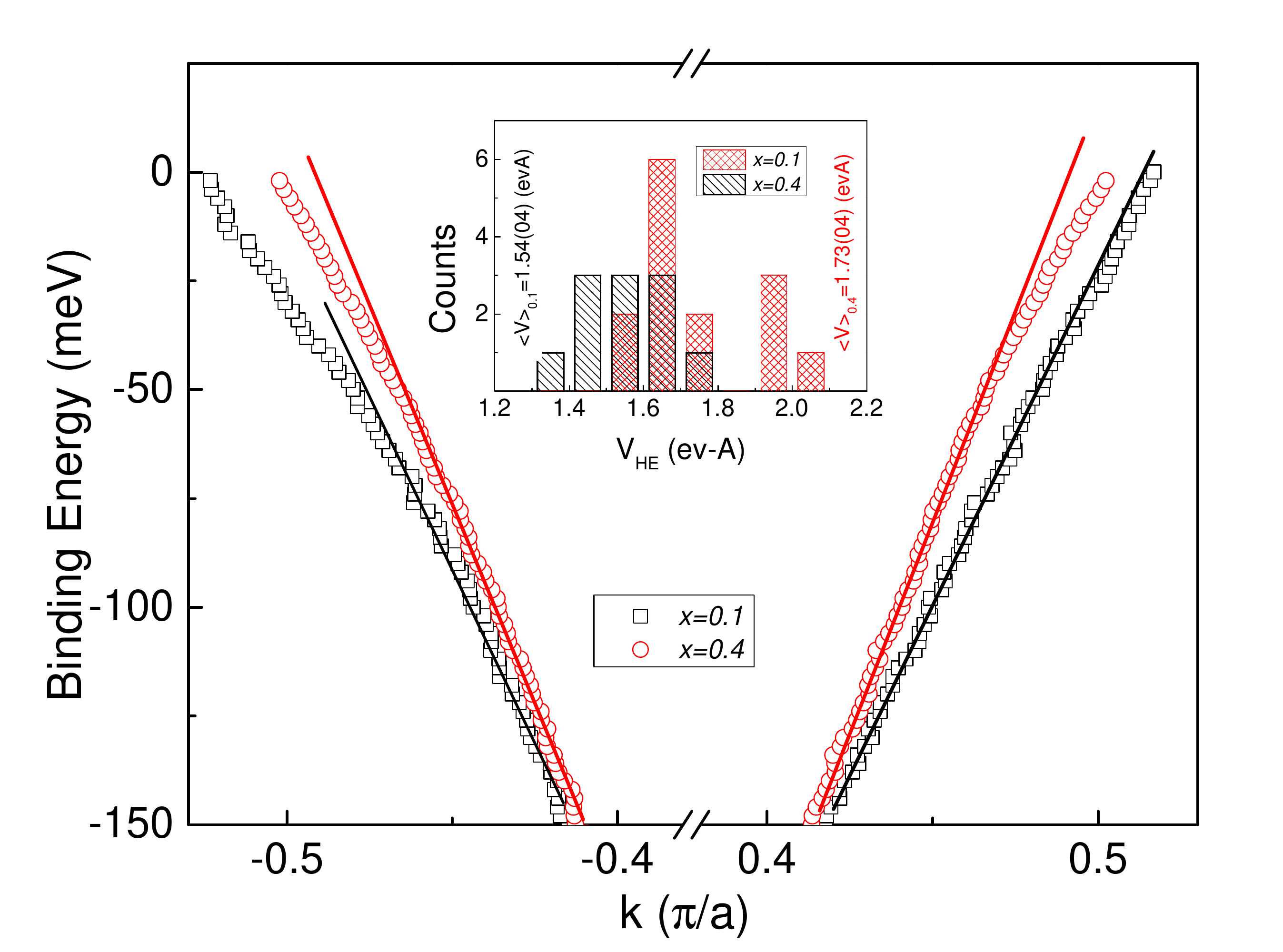}
\end{center}
\caption{ Main: The MDC peak position, extracted from the nodal dispersion
measured along $\Gamma -Y$, as a function of $k$ for $x=0.1$ (black squares)
and $x=0.4$ (red circles). Note the axis breaker. The solid lines are a
linear fit to the data in the $-150<E<-50$ meV range from which the high
energy velocities $V_{HE}$ are extracted. Inset: A histogram of high energy
velocities obtained from a series of CLBLCO samples with $x=0.1$ (black,
filled) and $x=0.4$ (red, crossed). }
\label{Velocite}
\end{figure}

Last but not least, we compare the nodal velocity between families. This
study was performed on six $x=0.1$ and seven $x=0.4$ crystals. The
dispersion in the nodal direction, previously described in Fig~\ref{Fig1}%
(e,h), was measured for each sample in two branches with high statistics.
After an orientation procedure, which is described in the supplementary
material, the peak positions in the MDC of each measured dispersion was
extracted and plotted as a function of binding energy. Exemplary dispersions
of two samples are shown in Fig.~\ref{Velocite}. An axis breaker is used in
order to show the two branches. The breaker emphasizes the differences
between $k_{f}$ of the two samples, which in fact is very small. Two
different linear regimes are observed. The first regime involves low
energies close to $E_{f}$, between $-50<E<0$ meV. The second regime
corresponds to high energies where $-150<E<-50$ meV. The transition between
these regimes is known as the kink and involves the electrons dispersion
re-normalized due to correlations~\cite{Sato-Renormalization} or coupling
between electrons and low energy bosonic degrees of freedom \cite%
{LanzaraNature01}. The slope of the dispersion $\partial E/\partial k$
provides the velocity in the low ($v_{F}$) and high-energy (${v}_{HE}$)
regimes. The results are similar to other overdoped materials \cite%
{ZhouNature03}.

We did not find differences with statistical significance in $v_{F}$ between
samples with different $x$. This is in agreement with previous work \cite%
{ZhouNature03}. As for ${v}_{HE}$, the results are summarized in the inset
of Fig.~\ref{Velocite} as histograms. For the $x=0.1$ family, the average
high-energy velocity is $\left\langle {V}_{HE}^{0.1}\right\rangle =1.53(04)$
eV$\mathring{A}$, while for the $x=0.4$ it is $\left\langle {V}%
_{HE}^{0.4}\right\rangle =1.73(04)$ eV$\mathring{A}$. Despite the velocity
distribution overlap, the average velocities differ by 3.5$\sigma $, and
hence are statistically different with 99.5\% confidence. Using these
velocities we can now calculate all the TB coefficients by $\partial
E/\partial k=4at\sin (k_{f}^{node}a)$. The unit cell parameter $a=3.91$~\AA\ %
is nearly family independent~\cite{OferPRB08}. The coefficient are presented
in the supplementary material and are in agreement with previously published
values \cite{Bansil-TB}\cite{NormanTBBSSCO}.

From the data presented, we can draw several conclusions. First, we discuss
the ratio of velocities ${\left( {\left\langle {{V_{HE}^{0.4}}}\right\rangle
/}\left\langle {{V_{HE}^{0.1}}}\right\rangle \right) ^{2}}\simeq 1.26\pm 0.08
$. Despite the large error-bar, this ratio is very close to that expected
from the ratio of the super-exchange $J$ between families. This ratio is
given by $J(0.4)/J(0.1)=T_{c}^{max}(0.4)/T_{c}^{max}(0.1)\simeq 1.4$ (see
Fig.~\ref{Fig1}). Therefore, the $J\propto t^{2}$ relation is obeyed, and $%
T_{c}^{max}$ depends on orbital overlaps even when the measurements are done
in the doped phase.

However, ARPES measurements do not necessarily represent the bulk properties. For
example, the buckling angle might change close to the surface. Nevertheless,
if such a thing happens in CLBLCO, it might affect both families equally.
The fact that the ratio of $J$ measured magnetically agrees with the ratio
of $t^{2}$ measure by ARPES supports this notion.

Second, we discuss the gap. There are three possible scenarios that explain
the difference in the gap size: I) A scenario where disorder leads to
broadening of the band structure features in $x=0.1$ which hide the gap.
However, high-resolution powder x-ray diffraction \cite{Xray} and NMR
experiments \cite{KerenNJP09} indicate that $x=0.1$ samples are more ordered
than $x=0.4$ ones. II) A scenario where $\Delta $ exist only below $T_{c}$.
It could be that in our experiment the surface of the $x=0.4$ sample is
below $T_{c}$, but the $x=0.1$ surface, is not since its $T_{c}$ is lower.
In this case only the $x=0.4$ sample will show a gap. The problem with this
scenario is the observation of a Fermi arc in $x=0.4$, which does not exist
below $T_{c}$ in any other cuprate. III) A scenario where both samples are
above $T_{c}$, but there is an intrinsic difference in their gap size. The
problem here is again that in other materials there is no gap above $T_{c}$
in extreme overdoped samples \cite{Chatterjee}. Further experiments are
needed to clarify this point.

In conclusion, we present the first ARPES data from CLBLCO. We find that the
surface doping is independent of the bulk doping or the Ca to Ba ratio. We
also demonstrate that the gap can be measured in this system. The hopping
parameter $t$ is larger for $x=0.4$ than for $x=0.1$ in the over-doped
sides. This suggests that $T_{c}^{max}$ is correlated with electron-orbital
overlaps on neighboring sites.

This research was supported by the Israeli Science Foundation (ISF) and the joint German-Israeli DIP Project.

\section{Supplementary Material for Linking Dynamic and Thermodynamic
Properties of Cuprates; an ARPES study of (CaLa)(BaLa)$_{2}$Cu$_{3}$O$_{y}$.}

\subsection{Gap Angular Dependence}

\begin{figure}[htbp]
\begin{center}
\includegraphics[trim=2cm 0cm 0cm
0cm,clip=true,height=6cm]{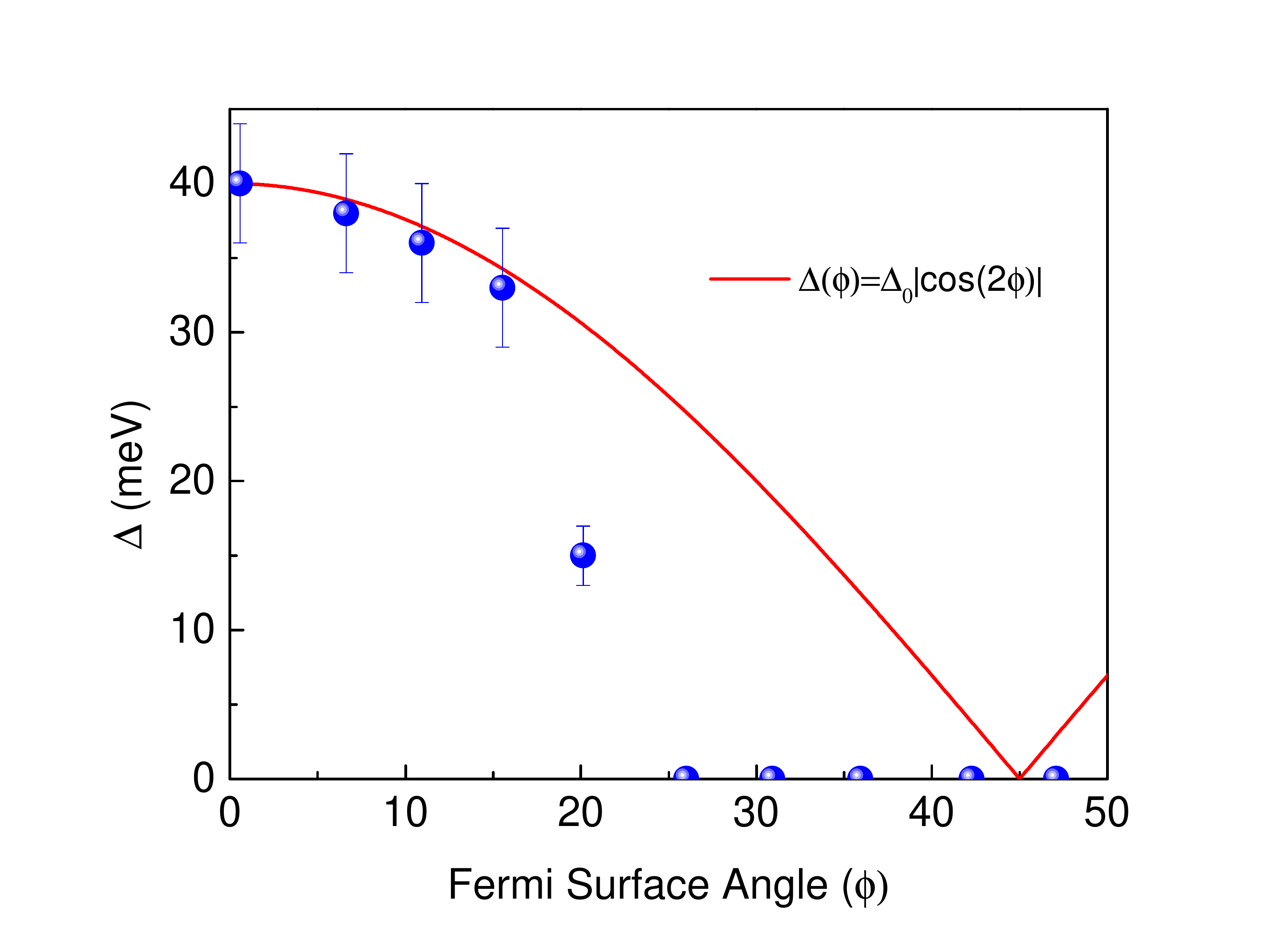}
\end{center}
\caption{ Gap size as a function of Fermi surface angle $\protect\phi $ for
the $x=0.4$ sample. The red line is the d-wave gap function, $\Delta (%
\protect\phi )=\Delta _{0}|\cos (2\protect\phi )|$, with $\Delta _{0}=40$
meV.}
\label{GapvsAngle}
\end{figure}

In Figure~\ref{GapvsAngle} we plot the gap size as a function of Fermi
surface angle for the $x=0.4$ sample. The value of gap, extracted from the
symmetrized EDC's shown in the article, is half of the peak-to-peak distance
or the change in slope of the EDC.

\subsection{Nodal Cut Measurement}

\begin{figure*}[htbp]
\begin{center}
\includegraphics[trim=0cm 0cm 0cm
0cm,clip=true,width=15cm]{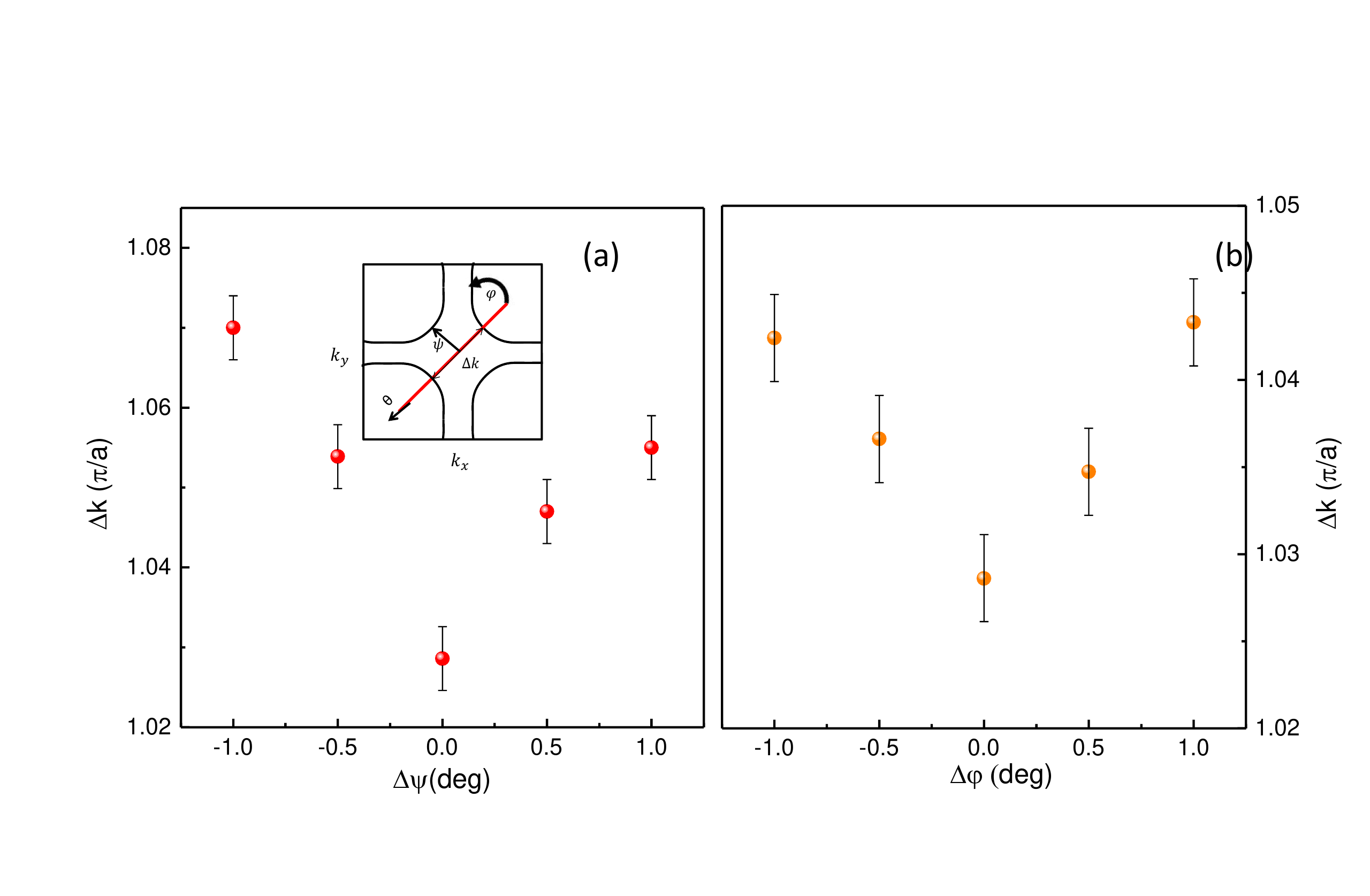}
\end{center}
\caption{ (a) The nodal distance $\Delta k$ as a function of the change in
angle $\protect\psi $, and (b) as a function of the change in angle $\protect%
\varphi $. The inset shows an illustration of a nodal cut upon a FS, with
the definition and action of $\protect\theta ,\protect\varphi $ and $\protect%
\psi $ angles used in the experiment}
\label{NNProcedure}
\end{figure*}

Because of the importance of proper orientation for the determination of the
nodal velocity, we developed an alignment protocol. First, we map the
complete FS of each sample in the first BZ. We define three angles, which
can be manipulated, as shown in the inset of Fig.~\ref{NNProcedure}(a). $%
\theta =0$ , $\varphi =45$, and $\psi $ $=0$ define a nodal cut. The red
strait line represents the analyzer opening. Second, the angle $\theta $ is
adjusted to give a symmetric spectrum. Third, we performed measurements with
an intentional shifts $\Delta \psi $ and $\Delta \varphi $ angle to ensure
truly perfect alignment. From each measurement we extracted the $k-$space
distance $\Delta k(\Delta \psi ,\Delta \varphi )$ between the two Fermi
points. Figure~\ref{NNProcedure}(a) presents $\Delta k(\Delta \psi ,0)$ for $%
\psi $ variations in steps of 0.5 a degrees. Figure ~\ref{NNProcedure}(b)
depicts $\Delta k(0,\Delta \varphi )$ for $\varphi $ rotations again in
steps of 0.5 degrees. Due to the geometry of the FS the nodal distance
should be shortest when the alignment is perfect. Indeed, in both cases, the
shortest distance was measured when $\Delta \psi =\Delta \varphi =0$. This
procedure was repeated for each and every measured sample.

\subsection{Tight Binding Parameters}

The tight binding parameters for CLBLCO extracted from the Fermi surface and
nodal velocity are given in the table.
\begin{table}[tbph]
\centering
\begin{tabular}{lllr}
\hline
i & $t_{i}^{x=0.1}$ & $t_{i}^{x=0.4}$ & $\eta_i(\mathbf{k})$ \\ \hline
0~~~~ & 0.134 & 0.152 & 1 \\
1~~~~ & 0.110 & 0.125 & $- 2\left[ {\cos \left( {{k_x}a} \right) + \cos
\left( {{k_y}a} \right)} \right]$ \\
2~~~~ & -0.032 & -0.036 & $- 4\left[ {\cos \left( {{k_x}a} \right)\cos
\left( {{k_y}a} \right)} \right]$ \\
3~~~~ & 0.016 & 0.018 & $- 2\left[ {\cos \left( {2{k_x}a} \right) + \cos
\left( {2{k_y}a} \right)} \right]$ \\ \hline
\end{tabular}%
\caption{Tight-binding coefficients and basis functions used to fit the
experimental data. The second column lists the coefficient of each term in
eV for the $x=0.1$ and $x=0.4$ samples, following the convention: $\protect%
\varepsilon \left( \mathbf{k}\right) =\sum {{t_{i}}{\protect\eta _{i}}\left(
\mathbf{k}\right) }$}
\label{table1}
\end{table}

\end{document}